\documentclass[12pt]{article}
\usepackage{bbm,latexsym}
\usepackage{amssymb,amsmath}
\usepackage{epsfig}
\newcommand{\be}{\begin{equation}}
\newcommand{\ee}{\end{equation}}
\newcommand{\ba}{\begin{eqnarray}}
\newcommand{\ea}{\end{eqnarray}}
\newcommand{\bs}{\begin{subequations}}
\newcommand{\es}{\end{subequations}}

\newcommand{\mnu}{\mathcal{M}_\nu}

\newcommand{\ds}{\Delta m^2_{21}}
\newcommand{\da}{\Delta m^2_{31}}

\begin{document}
\renewcommand{\thefootnote}{\fnsymbol{footnote}}

\title{
\normalsize \hfill UWThPh-2012-28 \\[10mm]
\LARGE Two-parameter neutrino mass matrices \\
with two texture zeros}

\author{
W.~Grimus\thanks{E-mail: walter.grimus@univie.ac.at} \
and P.O.~Ludl\thanks{E-mail: patrick.ludl@univie.ac.at}
\\[5mm]
\small University of Vienna, Faculty of Physics \\
\small Boltzmanngasse 5, A--1090 Vienna, Austria
}

\date{14 February 2013}

\maketitle

\begin{abstract}
We reanalyse Majorana-neutrino mass matrices $\mnu$ with two texture
zeros, by searching for viable hybrid textures in which the non-zero
matrix elements of $\mnu$ have simple ratios. 
Referring to the classification scheme of Frampton, Glashow
and Marfatia, we find that the mass matrix denoted by A1 allows the ratios 
$(\mnu)_{\mu\mu} : (\mnu)_{\tau\tau} = 1:1$ and 
$(\mnu)_{e\tau} : (\mnu)_{\mu\tau} = 1:2$. There are analogous 
ratios for texture A2. With these two hybrid textures, one
obtains, for instance, good agreement with the data if one computes
the three mixing angles in terms of the experimentally determined 
mass-squared differences $\da$ and $\ds$. We could not find 
viable hybrid textures based on mass matrices different from those of
cases A1 and A2.

\end{abstract}

\newpage

\renewcommand{\thefootnote}{\arabic{footnote}}

\section{Introduction}

With the recent results of the Double Chooz,
Daya Bay and RENO Collaborations~\cite{daya-reno}
a non-zero reactor mixing angle $\theta_{13}$ has been established. 
Since the value of 
$\theta_{13}$~\cite{daya-reno,forero,fogli,gonzalez} 
is rather large, neutrino mixing may not just be a perturbation of 
tri-bimaximal mixing~\cite{HPS}. Therefore, 
in model building~\cite{review} one might dispense with seeking for
contrived 
models which have tri-bimaximal mixing at leading order.

Therefore, 
it is interesting to revisit  
the most simple type of models, namely those whose predictivity is based on
texture zeros. Such texture zeros, which can in principle
always be imposed by Abelian symmetries at the expense of
an enlarged scalar sector~\cite{GJLT}, induce relations between
neutrino mass ratios and the mixing parameters.
It was already shown some time ago 
by Frampton, Glashow and Marfatia 
that, 
in the basis where the charged-lepton mass matrix is diagonal, 
there are only seven viable Majorana neutrino
mass matrices with two texture zeros~\cite{FGM}.
There are many investigations of two texture zeros in 
$\mnu$---see~\cite{fukuyama,xing,merle,dev,lashin,fritzsch,ludl,meloni}
and references therein, 
but in this paper we concentrate on so-called hybrid
textures~\cite{kaneko} which have a richer structure:  Apart from two
elements being zero, there are non-zero elements of $\mnu$ whose ratio
is one or another small integer. 
The motivation for considering hybrid textures is purely
phenomenological; if one finds a simple $\mnu$ which is able to
reproduce the experimentally established neutrino masses and lepton
mixing parameters, one might obtain a clue for an underlying symmetry.
In order to find viable hybrid
textures, we use a numerical method to estimate the elements
of $\mnu$~\cite{merle,meloni}. Then for each
$\mnu$ we make a hypothesis which ratios could be simple and accept or
refute the hypothesis on the basis of a $\chi^2$-analysis. 

In this investigation, 
the textures denoted by 
\begin{equation}
\mbox{A1:} \quad
\mnu \sim \left( \begin{array}{ccc}
0 & 0 & \times \\ 
0 & \times & \times \\
\times & \times & \times 
\end{array}
\right), \quad
\mbox{A2:} \quad
\mnu \sim \left( \begin{array}{ccc}
0 & \times & 0 \\ 
\times & \times & \times \\
0 & \times & \times 
\end{array}
\right)
\end{equation}
in the classification of the seven viable textures by
Frampton, Glashow and Marfatia~\cite{FGM}
will play a prominent role; in this equation, non-zero elements of
$\mnu$ are indicated by a cross.
We will find numerically that these textures 
can be endowed with a viable hybrid structure.
As a side remark, we note that the textures A1 and A2 allow only the normal
ordering of neutrino masses~\cite{FGM}. 
Moreover, neutrinoless $\beta\beta$-decay
by Majorana neutrino exchange is forbidden because the corresponding
effective neutrino mass is given by 
$\left( \mnu \right)_{ee} = 0$. 
Therefore, if such a decay is discovered, it must proceed via an
alternative mechanism.

The paper is organized as follows. In section~2 we describe 
our numerical method for estimating the elements of $\mnu$ and 
reanalyse two texture zeros.
In section~3 we use these results for 
guessing possible hybrid textures. As anticipated above, we find
two viable ones, which we 
denote by $\bar{\mathrm{A}}1$ and $\bar{\mathrm{A}}2$, 
based on textures A1 and A2, respectively. The remainder of 
section~3 is devoted to the discussion of 
$\bar{\mathrm{A}}1$ and $\bar{\mathrm{A}}2$.
The conclusions are presented in 
section~4.

\section{Two texture zeros and predictions for $|(\mathcal{M}_{\nu})_{ij}|$}

In~\cite{meloni} a $\chi^2$-analysis was performed in order to
investigate how well the seven allowed types of two texture zeros fit
with the available neutrino data. In this approach, 
a $\chi^2$-function incorporates the
experimental data and texture zeros have to be imposed. 
Here we will pursue an alternative approach. 
We define the function
\begin{equation}\label{F}
F_{ijkl}(x):=|(\mathcal{M}_{\nu})_{ij}(x)| +
|(\mathcal{M}_{\nu})_{kl}(x)|\quad\quad\quad (i,j,k,l=e,\mu,\tau), 
\end{equation}
where
\begin{equation}
x=(m_0,\,\Delta m_{21}^2,\, \Delta m_{31}^2,\, s_{12}^2,\, s_{23}^2,\,
s_{13}^2,\,\delta,\,\rho,\,\sigma). 
\end{equation}
In the parameter set $x$, 
$m_0$ is the lightest neutrino mass, 
$\Delta m_{ab}^2$ are the mass squared differences, 
$s_{cd}^2=\mathrm{sin}^2\theta_{cd}$ are the sines
squared of the mixing angles,
$\delta$ is the Dirac CP phase
and $\rho,\,\sigma$ are the two Majorana phases. 
For the lepton mixing matrix we use the parameterization of~\cite{rpp}.
Evidently, minimizing $F_{ijkl}$, taking into account the
experimentally allowed ranges of the parameters in 
$x$, should unveil the allowed two texture zeros. In order to
minimize $F_{ijkl}$ numerically, 
we make use of the Nelder--Mead algorithm (downhill simplex
method)~\cite{nelder-mead}. 
Since this algorithm is only capable of searching the whole parameter
space $\mathbbm{R}^9$, we add the function 
\begin{equation}
\Pi_D(x):=
\begin{cases}
0\, \mbox{eV}, & \text{for }x\in D\\
10^6\, \mbox{eV}, & \text{for }x\not\in D
\end{cases}
\end{equation}
to $F_{ijkl}(x)$, in order to specify the allowed domain $D$ for
$x$. To improve the chance 
of finding a good minimum, we started with 50000 different random
start simplices and also performed perturbations around 
good candidates for minima.

The domain $D$ of $x$ was chosen as follows. For $\Delta m_{ab}^2$,
$s_{cd}^2$ and $\delta$ we used the $1\sigma$ and $3\sigma$ ranges 
provided by the global fits of neutrino oscillation
data~\cite{forero,fogli}. The Majorana phases $\rho$ and $\sigma$ were
allowed to vary between $0$ and $2\pi$. 
Since most cosmological constraints on
the sum of the neutrino masses~\cite{rpp} are of 
the order of $\sum_{\nu} m_{\nu}< 1\,\mathrm{eV}$, we allowed the
lightest neutrino mass $m_0$ to vary between zero and 
$0.3$\,eV.

We minimized all 15 independent functions $F_{ijkl}$, using 
the $3\sigma$ and $1\sigma$ ranges provided 
by the global fits of
Forero~\textit{et al.}~\cite{forero} (versions 2 and 3) and
Fogli~\textit{et al.}~\cite{fogli}.\footnote{The fit results
  of~\cite{gonzalez} resemble those of Forero~\textit{et al.}
  (version~3). Therefore, we do not use them in the following.} 
The results for the $3\sigma$ range of Forero~\textit{et al.}
(version~3) and Fogli~\textit{et al.} are shown 
in table~\ref{t1}. The results for the corresponding $1\sigma$ ranges
can be found in table~\ref{t3}. 

As can be read off from tables~\ref{t1} and~\ref{t3}, due to the
finite accuracy of the numerical procedure, 
none of the functions $F_{ijkl}$ is found to have a minimum at exactly
zero. However, the allowed texture zeros 
are readily distinguished from the disfavoured ones. For the allowed
texture zeros, which are indicated in the 
tables by using boldface letters, we find numerical minima of the
order of $10^{-9} \div 10^{-8}$\,eV, 
while all other minima are at least four orders of magnitude larger.

Let us first discuss the results for the $3\sigma$ range---see
table~\ref{t1}. At $3\sigma$ the data of Forero~\textit{et al.} (both 
versions~2 and~3) allow precisely those two texture zeros, which are
also allowed following the original analysis~\cite{FGM}, 
namely A1--C for the normal and B1--C for the inverted neutrino mass
spectrum. The same holds for the data of Fogli~\textit{et al.}

This picture drastically changes, when we turn to the $1\sigma$
analysis---see table~\ref{t3}. The data of 
Forero~\textit{et al.} (version 3) now exclude B2 (inverted), B4
(inverted) and C (normal). The data of 
Fogli~\textit{et al.} allow only A1 (normal) at the $1\sigma$ level.
The older global fit by Forero~\textit{et al.} (version~2) allows all
``classical'' two texture zeros 
A1--C (normal) and B1--C (inverted) also at the $1\sigma$ level.

\begin{table}[t]
\begin{center}
\begin{tabular}{cccccccccc}
   & $ij$ $kl$ & \multicolumn{2}{c}{Forero \textit{et al.} (version 3)} & \multicolumn{2}{c}{Fogli \textit{et al.}}\\
   &           & normal & inverted & normal & inverted\\
A1 & $ee$ $e\mu$ & \boldmath{$5.51\times 10^{-9}$} & $3.47\times 10^{-2}$ & \boldmath{$5.09\times 10^{-9}$} & $3.57\times 10^{-2}$\\
A2 & $ee$ $e\tau$& \boldmath{$6.61\times 10^{-9}$} & $3.63\times 10^{-2}$ & \boldmath{$5.72\times 10^{-9}$} & $3.57\times 10^{-2}$\\ 
   & $ee$ $\mu\mu$ & $1.20\times 10^{-2}$ & $1.14\times 10^{-2}$ & $1.01\times 10^{-2}$ & $1.24\times 10^{-2}$\\
   & $ee$ $\mu\tau$ & $1.85\times 10^{-2}$ & $1.15\times 10^{-2}$ & $1.82\times 10^{-2}$ & $1.25\times 10^{-2}$\\
   & $ee$ $\tau\tau$ & $9.86\times 10^{-3}$ & $1.14\times 10^{-2}$ & $1.17\times 10^{-2}$ & $1.24\times 10^{-2}$\\
   & $e\mu$ $e\tau$ & $6.07\times 10^{-4}$ & $5.96\times 10^{-4}$ & $5.84\times 10^{-4}$ & $6.00\times 10^{-4}$\\
B3 & $e\mu$ $\mu\mu$ & \boldmath{$9.63\times 10^{-9}$} & \boldmath{$5.73\times 10^{-9}$} & \boldmath{$9.37\times 10^{-9}$} & \boldmath{$4.60\times 10^{-9}$}\\
   & $e\mu$ $\mu\tau$ & $1.82\times 10^{-3}$ & $1.83\times 10^{-3}$ & $1.71\times 10^{-3}$ & $1.79\times 10^{-3}$\\
B2 & $e\mu$ $\tau\tau$ & \boldmath{$6.86\times 10^{-9}$} & \boldmath{$6.47\times 10^{-9}$} & \boldmath{$7.60\times 10^{-9}$} & \boldmath{$9.31\times 10^{-9}$}\\
B1 & $e\tau$ $\mu\mu$ & \boldmath{$1.03\times 10^{-8}$} & \boldmath{$9.90\times 10^{-9}$} & \boldmath{$8.85\times 10^{-9}$} & \boldmath{$9.96\times 10^{-9}$}\\
   & $e\tau$ $\mu\tau$ & $1.77\times 10^{-3}$ & $1.79\times 10^{-3}$ & $1.75\times 10^{-3}$ & $1.79\times 10^{-3}$\\
B4 & $e\tau$ $\tau\tau$ & \boldmath{$9.94\times 10^{-9}$} & \boldmath{$8.89\times 10^{-9}$} & \boldmath{$5.11\times 10^{-9}$} & \boldmath{$9.35\times 10^{-9}$}\\
   & $\mu\mu$ $\mu\tau$ & $3.64\times 10^{-2}$ & $6.24\times 10^{-4}$ & $3.35\times 10^{-2}$ & $1.76\times 10^{-3}$\\
C  & $\mu\mu$ $\tau\tau$ & \boldmath{$6.05\times 10^{-9}$} & \boldmath{$5.97\times 10^{-9}$} & \boldmath{$7.65\times 10^{-9}$} & \boldmath{$8.00\times 10^{-9}$}\\
   & $\mu\tau$ $\tau\tau$ & $3.31\times 10^{-2}$ & $1.46\times 10^{-3}$ & $3.60\times 10^{-2}$ & $1.71\times 10^{-3}$\\
\end{tabular}
\end{center}
\caption{
  Minimal values of the functions $F_{ijkl}$ for the normal and the
  inverted ordering of the neutrino masses. The results are obtained
  by allowing the input parameters taken from~\cite{forero} and~\cite{fogli},
  respectively, to vary in the $3\sigma$ range. The entries are in units of
  eV. For viable texture zeros (entries in boldface) the minima are typically
  of the order of $10^{-9} \div 10^{-8}$, 
  while the non-viable textures have minima which are at least four orders of
  magnitude larger. In the left column, the notation for the ``acceptable'' two
  texture zeros according to~\cite{FGM} is found. \label{t1}
}
\end{table}
\begin{table}
\begin{center}
\begin{tabular}{cccccccccc}
   & $ij$ $kl$ & \multicolumn{2}{c}{Forero \textit{et al.} (version 3)} & \multicolumn{2}{c}{Fogli \textit{et al.}}\\
   &           & normal & inverted & normal & inverted\\
A1 & $ee$ $e\mu$ & \boldmath{$9.01\times 10^{-9}$} & $4.05\times 10^{-2}$ & \boldmath{$2.06\times 10^{-8}$} & $4.71\times 10^{-2}$\\
A2 & $ee$ $e\tau$& \boldmath{$1.61\times 10^{-8}$} & $4.70\times 10^{-2}$ & $1.60\times 10^{-3}$ & $4.34\times 10^{-2}$\\ 
   & $ee$ $\mu\mu$ & $1.51\times 10^{-2}$ & $1.52\times 10^{-2}$ & $1.29\times 10^{-2}$ & $1.97\times 10^{-2}$\\
   & $ee$ $\mu\tau$ & $2.04\times 10^{-2}$ & $1.56\times 10^{-2}$ & $2.02\times 10^{-2}$ & $1.67\times 10^{-2}$\\
   & $ee$ $\tau\tau$ & $1.32\times 10^{-2}$ & $1.54\times 10^{-2}$ & $2.60\times 10^{-2}$ & $1.62\times 10^{-2}$\\
   & $e\mu$ $e\tau$ & $7.46\times 10^{-4}$ & $7.44\times 10^{-4}$ & $7.48\times 10^{-4}$ & $7.57\times 10^{-4}$\\
B3 & $e\mu$ $\mu\mu$ & \boldmath{$1.32\times 10^{-8}$} & \boldmath{$1.08\times 10^{-8}$} & $4.14\times 10^{-3}$ & $1.93\times 10^{-2}$\\
   & $e\mu$ $\mu\tau$ & $1.99\times 10^{-3}$ & $2.02\times 10^{-3}$ & $1.91\times 10^{-3}$ & $1.96\times 10^{-3}$\\
B2 & $e\mu$ $\tau\tau$ & \boldmath{$6.49\times 10^{-9}$} & $1.74\times 10^{-2}$ & $2.38\times 10^{-2}$ & $8.37\times 10^{-4}$\\
B1 & $e\tau$ $\mu\mu$ & \boldmath{$7.70\times 10^{-9}$} & \boldmath{$7.01\times 10^{-9}$} & $3.29\times 10^{-3}$ & $1.86\times 10^{-2}$\\
   & $e\tau$ $\mu\tau$ & $1.95\times 10^{-3}$ & $1.97\times 10^{-3}$ & $2.14\times 10^{-3}$ & $2.02\times 10^{-3}$\\
B4 & $e\tau$ $\tau\tau$ & \boldmath{$6.24\times 10^{-9}$} & $1.84\times 10^{-2}$ & $2.35\times 10^{-2}$ & $1.23\times 10^{-3}$\\
   & $\mu\mu$ $\mu\tau$ & $4.11\times 10^{-2}$ & $5.08\times 10^{-3}$ & $3.79\times 10^{-2}$ & $1.90\times 10^{-2}$\\
C  & $\mu\mu$ $\tau\tau$ & $1.16\times 10^{-2}$ & \boldmath{$4.21\times 10^{-9}$} & $1.88\times 10^{-2}$ & $5.56\times 10^{-3}$\\
   & $\mu\tau$ $\tau\tau$ & $3.85\times 10^{-2}$ & $1.05\times 10^{-3}$ & $5.01\times 10^{-2}$ & $5.95\times 10^{-3}$
\end{tabular}
\end{center}
\caption{
  Minimal values of the functions $F_{ijkl}$ for the normal and the
  inverted ordering of the neutrino masses. The input parameters are allowed
  to vary in the $1\sigma$ range. For further information \textit{cf.}
  table~\ref{t1}. \label{t3}
}
\end{table}

\section{Two simple hybrid textures for $\mathcal{M}_{\nu}$}

As an additional gain of the analysis presented in section~2, we can calculate
the absolute values of the elements of $\mathcal{M}_{\nu}$ at the minima
of $F_{ijkl}$. 
In this way we find that some of the absolute values
$|(\mathcal{M}_{\nu})_{ij}|$ are approximately equal for textures of
type~B and~C in the classification of~\cite{FGM}, namely
$|(\mathcal{M}_{\nu})_{ee}| \approx |(\mathcal{M}_{\nu})_{\mu\tau}|$. 
It is interesting to note that this
approximate equality is more pronounced in version~2 of~\cite{forero}
than in version~3 or in~\cite{fogli}. Anyway, replacing the
approximate equality by an exact equality does not work for matrices
of type~B and~C. We have checked this by using a $\chi^2$-analysis and
have always obtained a very large $\chi^2$, mostly because one or two
mixing angles could not be reproduced.

It remains to discuss the textures of type~A. 
For definiteness we show the results for the $1\sigma$-analysis of
A1, using the data of Forero~\textit{et al.} (versions~2 and~3) 
and Fogli~\textit{et al.}:
\begin{eqnarray}
\label{A1-forero2}
\left(
\begin{array}{ccc}
2.10\times 10^{-9} & 5.08\times 10^{-9} & 1.20\times 10^{-2}\\
5.08\times 10^{-9} & 2.86\times 10^{-2} & 2.25\times 10^{-2}\\
1.20\times 10^{-2} & 2.25\times 10^{-2} & 2.60\times 10^{-2}
\end{array}
\right) && \text{(Forero~\textit{et al.}, version~2)}, \\
\label{A1-forero3}
\left(
\begin{array}{ccc}
5.65\times 10^{-9} & 3.36\times 10^{-9} & 1.12\times 10^{-2}\\
3.36\times 10^{-9} & 2.59\times 10^{-2} & 2.32\times 10^{-2}\\
1.12\times 10^{-2} & 2.32\times 10^{-2} & 2.70\times 10^{-2}
\end{array}
\right) && \text{(Forero~\textit{et al.}, version~3)}, \\
\label{A1-fogli}
\left(
\begin{array}{ccc}
1.70\times 10^{-8} & 3.62\times 10^{-9} & 1.07\times 10^{-2}\\
3.62\times 10^{-9} & 2.32\times 10^{-2} & 2.25\times 10^{-2}\\
1.07\times 10^{-2} & 2.25\times 10^{-2} & 2.94\times 10^{-2}
\end{array}
\right) && \text{(Fogli~\textit{et al.})}.
\end{eqnarray}
Looking at these matrices, we make the following observations. 
First of all, the texture zeros are represented by entries of the
order of $10^{-9} \div 10^{-8}$\,eV, while all other entries are of the
order of $10^{-2}$\,eV. Second, we see that 
$|(\mathcal{M}_{\nu})_{\mu\mu}| \approx |(\mathcal{M}_{\nu})_{\tau\tau}|$
and
$2|(\mathcal{M}_{\nu})_{e\tau}| \approx |(\mathcal{M}_{\nu})_{\mu\tau}|$
in equations~(\ref{A1-forero2}) and~(\ref{A1-forero3}), 
but these approximate relations are less pronounced with the data of 
Fogli~\textit{et al.}---see equation~(\ref{A1-fogli}). 
One finds analogous results for texture A2.
Contrary to textures of type~B and~C, 
replacing the two approximate equalities by exact equalities, we find
good to moderately good fits. This peculiar result conforms to
the statement in~\cite{meloni} that textures of type~A need
less finetuning to reproduce the oscillation parameters than those of
type~B and~C. Indeed, we find that changing the elements
of $\mnu$ slightly in order to achieve exact equalities works very
well for type~A, while for type~B and~C this procedure fails.

The bottom line of this discussion is that in the following we will 
consider the mass matrices
\begin{equation}\label{AB}
\bar{\mathrm{A}}1: \quad \mnu = 
\left( \begin{array}{ccc}
0 & 0 & a \\ 
0 & b & 2a \\
a & 2a & b
\end{array}
\right), \quad
\bar{\mathrm{A}}2: \quad \mnu = 
\left( \begin{array}{ccc}
0 & a & 0 \\ 
a & b & 2a \\
0 & 2a & b
\end{array}
\right).
\end{equation}
Actually, it suffices to discuss $\bar{\mathrm{A}}1$, since 
$\bar{\mathrm{A}}2$ emerges from $\bar{\mathrm{A}}1$ through the
exchange $\mu \leftrightarrow \tau$. Therefore, we can use the
following theorem~\cite{lashin,fritzsch}: 
Given two Majorana mass matrices $\mnu$ and
$\mnu'$, in the basis where the charged-lepton mass matrix is
diagonal, and 
\begin{equation}
S \mnu S = \mnu' \quad \mbox{with} \quad
S = 
\left( \begin{array}{ccc}
1 & 0 & 0 \\ 0 & 0 & 1 \\ 0 & 1 & 0 
\end{array} \right),
\end{equation}
then the mixing parameters of $\mnu'$ are related to those of $\mnu$ by
\begin{equation}\label{1-2}
\theta'_{12} = \theta_{12}, \quad
\theta'_{13} = \theta_{13}, \quad
\theta'_{23} = \frac{\pi}{2} - \theta_{23}, \quad 
\delta' = \pi + \delta.
\end{equation}

The matrices~(\ref{AB}) contain three parameters, the absolute values
of $a$ and $b$, and the relative phase. However, fitting these mass
matrices to the oscillation parameters gives a relative phase rather
close to zero or $\pi$. Therefore, we go one step further and assume
that $a$ and $b$ are both \emph{real}. Thus we end up with a
two-parameter scheme for neutrino masses and mixing.\footnote{In this
  context it is interesting to note that in equation~(24)
  of~\cite{araki} a one-parameter hybrid texture for $\mnu$ based on
  texture A1, which is \emph{not} a special case of
  equation~(\ref{AB}), has been proposed.}  

With real mass matrices~(\ref{AB}), their eigenvalues $\mu_j$ agree
with the neutrino masses $m_j$ apart from possible signs, \textit{i.e.}
$m_j = | \mu_j |$. Both cases 
$\bar{\mathrm{A}}1$ and $\bar{\mathrm{A}}2$ lead to the equations 
\begin{eqnarray}
2b & = & \mu_1 + \mu_2 + \mu_3, \label{2b} \\
b^2 - 5a^2 & = & \mu_1 \mu_2 + \mu_2 \mu_3 + \mu_3 \mu_1, \\
-a^2 b & = & \mu_1 \mu_2 \mu_3 \label{-a2b}.
\end{eqnarray}
Since there are three equations but only two real parameters, we find
the consistency condition
\begin{equation}\label{consistency}
\frac{1}{4} \left( \mu_1 + \mu_2 + \mu_3 \right)^2 + 
10\, \frac{\mu_1 \mu_2 \mu_3}{\mu_1 + \mu_2 + \mu_3} - 
\left( \mu_1 \mu_2 + \mu_2 \mu_3 + \mu_3 \mu_1 \right) = 0.
\end{equation}
By a rephasing of $\mnu$ we can always achieve $a > 0$
and $b > 0$, which entails, due to equations~(\ref{2b}) and~(\ref{-a2b}), 
\begin{equation}\label{ineq}
\mu_1 + \mu_2 + \mu_3 > 0 
\quad \mbox{and} \quad
\mu_1 \mu_2 \mu_3 < 0,
\end{equation}
respectively.

Since the textures A1 and A2 allow only the normal ordering of the
neutrino mass spectrum, the masses $m_2$ and $m_3$ can be expressed as
a function of the lightest mass $m_1$ and the mass-squared differences:
\begin{equation}\label{m2m3}
m_2 = \sqrt{m_1^2 + \ds}, \quad m_3 = \sqrt{m_1^2 + \da}.
\end{equation}
As a consequence, we can conceive equation~(\ref{consistency}) as
condition to determine $m_1$, and, therefore, also $m_2$ and $m_3$, 
as a function of the measured
mass-squared differences $\ds$ and $\da$. However, then we also obtain
$a$ and $b$ in terms of the mass-squared differences. All told, the
neutrino mass spectrum and the mixing angles are, in this line of
reasoning, functions of $\ds$ and $\da$. In the following,
we will discuss texture~$\bar{\mathrm{A}}1$. The results
corresponding to texture~$\bar{\mathrm{A}}2$ are easily obtained via
equation~(\ref{1-2}). 

In the light of the above paragraph, we use as input the best fit
values for $\ds$ and $\da$, in order to determine the neutrino masses
and the mixing angles. This input is 
$\ds = 7.62 \times 10^{-5}$\,eV$^2$ and 
$\da = 2.55 \times 10^{-3}$\,eV$^2$ from~\cite{forero} (version~3) and 
$\ds = 7.54 \times 10^{-5}$\,eV$^2$ and 
$\da = 2.47 \times 10^{-3}$\,eV$^2$ from~\cite{fogli}; in the latter
case we computed $\da$ via $\da = \Delta m^2 + \delta m^2/2$ where 
$\delta m^2 \equiv \ds$ and $\Delta m^2 \equiv m_3^2 - (m_1^2 + m_2^2)/2$.
First we discuss the sign ambiguities. Due to our sign convention, the
sign of $\mu_3$ is determined by the signs of $\mu_1$ and $\mu_2$ via
the second inequality in equation~(\ref{ineq}). It turns out that
equation~(\ref{consistency}) numerically admits only 
the signs such that $\mu_1 \mu_2 < 0$. Therefore, the two possibilities
which remain are $\mu_1 < 0$, $\mu_2 > 0$ and 
$\mu_1 > 0$, $\mu_2 < 0$. Moreover, for each of these two
possibilities the real solution of equation~(\ref{consistency}) is unique. 
After having obtained $m_1$ for each of the two sign possibilities, we
compute the mixing angles. It turns out
that the second sign
possibility is ruled out because it gives
a too big value for
both $s_{12}^2$ and $s_{13}^2$. The numerical results for the signs 
$\mu_1 < 0$, $\mu_2 > 0$ are presented in table~\ref{num-res}.
\begin{table}
\begin{center}
\begin{tabular}{c|cc}
& Forero & Fogli \\
& \textit{et al.} & \textit{et al.} \\\hline
$m_1$ & 0.62 & 0.60 \\
$m_2$ & 1.07 & 1.06 \\
$m_3$ & 5.09 & 5.01 \\
$\sin^2 \theta_{12}$ & 0.291 & 0.289 \\
$\sin^2 \theta_{23}$ & 0.473 & 0.473 \\
$\sin^2 \theta_{13}$ & 0.024 & 0.024 \\
$\cos\delta$ & $-1$ & $-1$ \\
$a$ & 1.10 & 1.08 \\
$b$ & 2.77 & 2.73 \\
\end{tabular}
\end{center}
\caption{Numerical results for texture $\bar{\mathrm{A}}1$, using the
  best-fit values for $\ds$ and $\da$ of 
  Forero \textit{et al.}~\cite{forero} (version~3) and 
  Fogli \textit{et al.}~\cite{fogli}, respectively, as input.
  The values of the neutrino masses and of the parameters $a$ and $b$ are
  given in units of $10^{-2}$\,eV.
\label{num-res}}
\end{table}

The mixing angles are practically insensitive to a variation of $\ds$
in the $3\sigma$ range. Under a variation of $\da$ in the $3\sigma$
range, $s^2_{23}$ and $s^2_{13}$ change only insignificantly. However, 
if $\da$ varies between
$2.2 \times 10^{-3}$\,eV$^2$ and $2.8 \times 10^{-3}$\,eV$^2$, which
covers the $3\sigma$ ranges of~\cite{forero} and~\cite{fogli},
$s^2_{12}$ varies between 0.273 and 0.301. Correspondingly, $m_1$ changes from 
$5.4 \times 10^{-3}$\,eV to $6.6 \times 10^{-3}$\,eV. 

The results for texture $\bar{\mathrm{A}}2$ are obtained by applying 
equation~(\ref{1-2}), leading to the replacements 
$\cos\delta = -1 \to +1$ and $s^2_{23} \to 1 - s^2_{23}$
in table~\ref{num-res}.
Consequently, in the case of texture
$\bar{\mathrm{A}}2$ the atmospheric mixing angle is in the second
octant. It is amusing to observe that the analysis of~\cite{fogli}
slightly prefers $\cos \delta = -1$ and, therefore, texture
$\bar{\mathrm{A}}1$, but this preference vanishes at the $2\sigma$
level.

\section{Conclusions}
In this paper we have assumed that the charged-lepton mass matrix is diagonal
and the Majorana neutrino mass matrix $\mnu$ has two texture zeros. 
We have revisited the 15 possibilities of two texture zeros and confirmed the
original seven viable texture zeros of~\cite{FGM} if we let the input
parameters of the function $F_{ijkl}$ of equation~(\ref{F}) vary individually
in the respective $3\sigma$ ranges provided 
by~\cite{forero} and~\cite{fogli}. However, if we permit
only the $1\sigma$ ranges, then in the case of 
Forero \textit{et al.}~\cite{forero}  
several of the previously viable textures cease to be so. Interestingly, if we
take the input from Fogli \textit{et al.}~\cite{fogli}, at $1\sigma$
only one viable texture, A1, remains. 
The reason for this is rooted mainly in the different
$1\sigma$ ranges of $s^2_{23}$ in these papers. However, since it is even
unclear in which octant the best fit value of $\theta_{23}$ lies, 
we should not put too much emphasis on that point. 

Searching for viable hybrid textures, we have found two such textures,
denoted by $\bar{\mathrm{A}}1$ and $\bar{\mathrm{A}}2$ in
equation~(\ref{AB}).\footnote{Hybrid textures with one texture zero,
  $(\mnu)_{ee} = 0$, have recently been investigated
  in~\cite{bentov}.}
Fitting these textures to the oscillation parameters suggests real
parameters $a$ and $b$. Thus we end up with two two-parameter textures
of the neutrino mass matrix. In other words, there are two parameters
for three mixing angles and two mass-squared differences. 
In order demonstrate the viability of 
$\bar{\mathrm{A}}1$ and $\bar{\mathrm{A}}2$,
we have adopted the strategy 
to compute the mixing angles from the fitted mass-squared
differences. The mixing angles
obtained in this way agree with the fit results
of~\cite{forero} at the $2\sigma$ level; this is also true
for~\cite{fogli}, with the exception of $s^2_{23}$ where the agreement is
at $3\sigma$.

Since the textures $\bar{\mathrm{A}}1$ and $\bar{\mathrm{A}}2$ are
special cases of the textures of type~A, 
they require normal ordering of the neutrino mass
spectrum. Numerically, it turns out that the spectrum is hierarchical
because $m_1/m_3 \simeq 1/8$.
Moreover, $\bar{\mathrm{A}}1$ predicts an atmospheric mixing angle in
the first octant and $\cos\delta = -1$, while for $\bar{\mathrm{A}}2$
$\theta_{23}$ lies in the second octant and $\cos\delta = +1$.
As a function of the mass-squared differences, our two hybrid textures
give $s^2_{23}$ fairly close to 0.5. This is in very good agreement
with version~2 of~\cite{forero}, 
but it is not completely consistent with the mean values
of $s^2_{23}$ in version~3 of the same authors and in~\cite{fogli}. 
But, as mentioned earlier, the
experimental situation with respect to $s^2_{23}$ is a bit ambiguous.

If a texture of type~A is realized in nature, then observation of
neutrinoless $\beta\beta$-decay would signal new physics. In the
case of our hybrid textures $\bar{\mathrm{A}}1$ and
$\bar{\mathrm{A}}2$, there would be no CP violation in neutrino
oscillations. This would be unfortunate, but at the moment the results
of~\cite{fogli}, although far from being significant, hint at 
$\delta \approx \pi$. As stressed in~\cite{meloni}, the textures 
A1 and A2 need the least finetuning of all two texture zeros and
they have no pronounced hierarchies in the matrix elements of $\mnu$. 
This is also borne out by the analysis of our hybrid textures---see 
table~\ref{num-res}. 

\vspace{5mm}

\noindent
\textbf{Acknowledgement:} This work is supported by the Austrian
Science Fund (FWF), Project No.\ P~24161-N16.

\end{document}